# Superluminal Near-field Dipole Electromagnetic Fields

**William D. Walker**
KTH-Visby
Cramérgatan 3
SE-621 57 Visby, Sweden
Email: bill@visby.kth.se

## 1 Introduction

The purpose of this paper is to present mathematical evidence that electromagnetic near-field waves and wave groups, generated by an oscillating electric dipole, propagate much faster than the speed of light as they are generated near the source, and reduce to the speed of light at about one wavelength from the source. The speed at which wave groups propagate (group speed) is shown to be the speed at which both modulated wave information and wave energy density propagate. Because of the similarity of the governing partial differential equations, two other physical systems (magnetic oscillating dipole, and gravitational radiating oscillating mass) are noted to have similar results.

## 2 Analysis of electric dipole

### 2.1 General solution

Numerous textbooks present solutions of the electromagnetic fields generated by an oscillating electric dipole[1,3]. One simple and elegant solution solves the inhomogeneous second order "superpotential" wave equation[1] (pp. 254 - 260). The electromagnetic fields can then be derived from the Hertz vector (Z).

**Figure 1.** Spherical coordinate system used in problem:

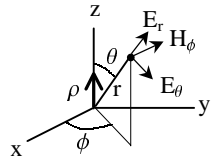

PDE (superpotential wave equation):
$$\Box Z = \frac{-\rho}{\varepsilon_o} \qquad (1)$$

Solution:
$$Z_R = \frac{\rho \, Cos(\theta) e^{i(kr)}}{4\pi\varepsilon_o r} \qquad Z_\theta = \frac{-\rho \, Sin(\theta) e^{i(kr)}}{4\pi\varepsilon_o r} \qquad Z_\phi = 0 \qquad (2)$$

Defining equations:
$$C = \nabla \times Z \qquad C_R = 0 \qquad C_\theta = 0 \qquad C_\phi = \frac{\rho \, Sin(\theta)}{4\pi\varepsilon_o r^2}[1 - ikr]e^{i(kr)} \qquad (3)$$

Field calculations:
$$E = \nabla \times C \qquad B = \frac{1}{c_o^2}\frac{\partial C}{\partial t} \qquad \text{where: } H = \frac{B}{\mu_o} \qquad (4)$$

Resultant electrical and magnetic field components for an oscillating electric dipole

$$E_r = \frac{\rho Cos(\theta)}{2\pi\varepsilon_o\, r^3}[1-i(kr)]e^{i(kr)} \qquad E_\theta = \frac{\rho Sin(\theta)}{4\pi\varepsilon_o\, r^3}\left[\{1-(kr)^2\}-i(kr)\right]e^{i(kr)} \qquad (5)$$

$$H_\varphi = \frac{\omega\rho Sin(\theta)}{4\pi\, r^2}[-kr-i]e^{i(kr)} \qquad (6)$$

It should be noted that this solution is only valid for distances (r) much greater than the dipole length ($d_o$). In the region next to the source (r ~ $d_o$) can not be modeled as a sinusoid: $Sin(\omega t)$. Instead it must be modeled as a sinusoid inside a dirac delta function: $\delta[r - d_o Sin(\omega t)]$. The solution to this hyper-near-field problem can be calculated using the Lienard-Wiehart potentials[13, 14, 18].

## 2.2  Lines of electric force analysis

Traditionally the electric lines of force can be determined from the relation that a line element (ds) crossed with the electric field is zero. The resulting partial differential equation can then be solved yielding the classical result.

Resultant Equation: $\quad \frac{\partial}{\partial r}(rC_\phi Sin\theta)dr + \frac{\partial}{\partial \theta}(rC_\phi Sin\theta)d\theta = 0 \qquad (7)$

Solution: $\quad \sqrt{1+\frac{1}{(kr)^2}}\; Sin^2\theta\; Cos[kr - Tan^{-1}(kr) - \omega t] = Const \qquad (8)$

A contour plot of this solution (Eq. 8) reveals the classical radiating oscillating electric dipole field pattern (Fig. 2). Careful examination of the pattern reveals that the wavelength of the generated fields are larger in the nearfield (a) and reduce to a constant wavelength after the fields have propagated about one wavelength from the source (b). The speed of the fields (phase speed = $c_{ph}$) near the source can then be concluded to propagate faster than the speed of light from the relation that wave speed ($c_{ph}$) is equal to the wavelength ($\lambda$) multiplied by the frequency (f), which is constant. $\qquad \lambda \cdot f = c_{ph} \qquad if\ \lambda \uparrow \therefore\ c_{ph} \uparrow$

**Figure 2.** Mathematica animation contour plot of above solution

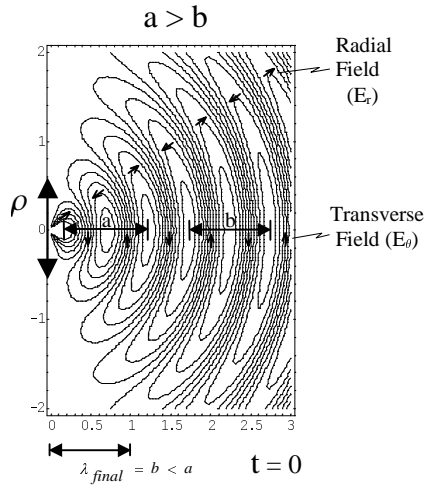

*Mathematica code used to generate animation*

```
L = 1; c = 3*10^8; f = c/L; T = 1/f; w = 2 N[Pi] f; k = w/c;
fn = Sqrt[1/(k r)^2 + 1] * Cos[Th]^2 * Cos[w t - k r + ArcTan[k r]];
r = Sqrt[x^2 + y^2]; Th = ArcTan[y/x];
Animate[ContourPlot[fn, {x, 0.01, 3}, {y, -2, 2}, PlotPoints -> 100],
{t, 0, 3*T}, ContourShading -> False,
Contours -> {-.9, -.7, -.5, -.3, -.1, .1, .3, .5, .7, .9},
AspectRatio -> 3/2]
```

The Mathematica code (Ver. 3.0) shown above generates 24 plots of the propagating electric field at different isolated moments in time. Mouse clicking any of the frames in Mathematica animates the plot, revealing that the ovals of constant electric field enlarge as they propagate away from the source (located at r = 0). It is also interesting to note that as the electric field lines are generated, some of the electric lines of force very near the dipole ($\sim \leq \lambda/10$ wavelength) appear to propagate only a short distance and then reverse and propagate back into the source.

**Figure 3.** Mathematica animation of contour plot of electric lines of force near source

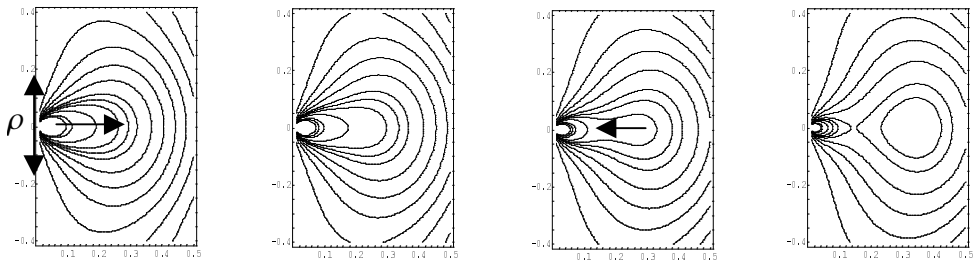

# 3  Analysis of phase speed and group speed

## 3.1  Phase calculation

The general form of the electromagnetic fields (ref Eq. 5,6) generated by a dipole is:

$$Field \propto (x+iy) \cdot e^{i[kr-\omega t]} \tag{9}$$

If the source is modeled as $Cos(\omega t)$, the resultant generated field is:

$$Field \propto Mag \cdot Cos[\{kr + ph\} - \omega t] = Mag \cdot Cos(\theta - \omega t) \tag{10}$$

$$\text{where: } Mag = \sqrt{x^2 + y^2}$$

It should be noted that the formula describing the phase is dependant on the quadrant of the complex vector.

$$\theta_1 = kr + Tan^{-1}\left(\frac{y}{x}\right) \qquad \theta_2 = kr - Cos^{-1}\left(\frac{x}{\sqrt{x^2+y^2}}\right) \tag{11}$$

## 3.2  Definition and calculation of wave phase speed

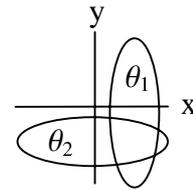

Phase speed can be defined as the speed at which a wave composed of one frequency propagates.

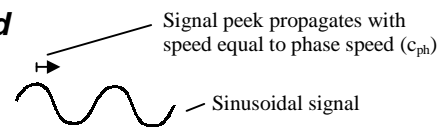

The phase speed ($c_{ph}$) of an oscillating field of the form $Sin(\omega t - kr)$, in which $k = k(\omega, r)$, can be determined by setting the phase part of the field to zero, differentiating the resultant equation, and solving for $\partial r / \partial t$.

$$\frac{\partial}{\partial t}(\omega t - kr) = 0 \qquad \therefore \omega - k\frac{\partial r}{\partial t} - r\frac{\partial k}{\partial r}\frac{\partial r}{\partial t} = 0 \tag{12}$$

Differentiating $\theta \equiv -kr$ with respect to r yields: $\quad \dfrac{\partial \theta}{\partial r} = -k - r\dfrac{\partial k}{\partial r}$ (13)

Combining these results and using $\omega = c_o k$ yields: $\quad c_{ph} = -\omega \left/ \dfrac{\partial \theta}{\partial r} \right. = \boxed{-c_o k \left/ \dfrac{\partial \theta}{\partial r}\right.}$ (14)

### 3.3 Definition and calculation of wave group speed

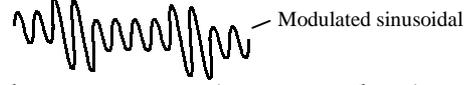

Peak of modulation part of signal propagates with speed equal to group speed ($c_g$)

Modulated sinusoidal

In some physical systems the wave phase speed is a function of frequency. In these systems when waves composed of different frequencies propagate, the wave group (wave envelope) propagates at a different speed (group speed) than the individual waves. The group speed is also known to be the speed at which wave energy and wave information propagate[6](pp.268-269), [10](p.123). The group speed of an oscillating field of the form $Sin(\omega t - kr)$, in which $k = k(\omega, r)$, can be calculated by considering two Fourier components of a wave group:

$$Sin(\omega_1 t - k_1 r) + Sin(\omega_2 t - k_2 r) = Sin(\Delta\omega t - \Delta k r)\, Sin(\omega t - kr) \tag{15}$$

in which: $\quad \Delta\omega = \dfrac{\omega_1 - \omega_2}{2}, \quad \Delta k = \dfrac{k_1 - k_2}{2}, \quad \omega = \dfrac{\omega_1 + \omega_2}{2}, \quad k = \dfrac{k_1 + k_2}{2}$

The group speed ($c_g$) can then be determined by setting the phase part of the modulation component of the field to zero, differentiating the resultant equation, and solving for $\partial r / \partial t$:

$$\frac{\partial}{\partial t}(\Delta\omega t - \Delta k r) = 0 \qquad \therefore \Delta\omega - \Delta k \frac{\partial r}{\partial t} - r\frac{\partial \Delta k}{\partial r}\frac{\partial r}{\partial t} = 0 \qquad \therefore c_g = \frac{\partial r}{\partial t} = \frac{\Delta\omega}{\Delta k + r\dfrac{\partial \Delta k}{\partial r}} \tag{16}$$

Differentiating $\Delta\theta \equiv -\Delta k r$ with respect to r yields: $\quad \dfrac{\partial \Delta\theta}{\partial r} = -\Delta k - r\dfrac{\partial \Delta k}{\partial r}$ (17)

Combining these results and using the relation $\omega = c_o k$ yields:

$$c_g = -\Delta\omega \left/ \frac{\partial \Delta\theta}{\partial r}\right. = -\left[\frac{\partial}{\partial r}\frac{\Delta\theta}{\Delta\omega}\right]^{-1}$$

$$\therefore c_g \underset{\substack{\lim \\ \frac{\Delta\theta}{\Delta\omega}\,small}}{=} -\left[\frac{\partial^2 \theta}{\partial r \partial \omega}\right]^{-1} = \boxed{-\frac{1}{c_o}\left[\frac{\partial^2 \theta}{\partial r \partial k}\right]^{-1}} \tag{18}$$

## 3.4 Radial electric field ($E_r$)

Applying the above phase and group speed relations (Eq. 14, 18) to the radial electrical field ($E_r$) component (Eq. 5) yields the following results:

**Figure 4**

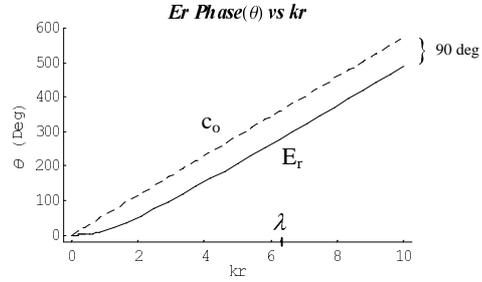

$$y = -kr \qquad x = 1 \qquad (19)$$

$$\theta = kr - Tan^{-1}(kr) \underset{kr \ll 1}{\approx} -\frac{1}{3}(kr)^3 \qquad (20)$$

$$c_{ph} = c_o\left(1 + \frac{1}{(kr)^2}\right) \underset{kr \ll 1}{\approx} \frac{c_o}{(kr)^2} \underset{kr \gg 1}{\approx} c_o \qquad (21)$$

$$c_g = \frac{c_o\left(1 + (kr)^2\right)^2}{3(kr)^2 + (kr)^4} \underset{kr \ll 1}{\approx} \frac{c_{ph}}{3} \underset{kr \gg 1}{\approx} c_o \qquad (22)$$

**Figure 5**

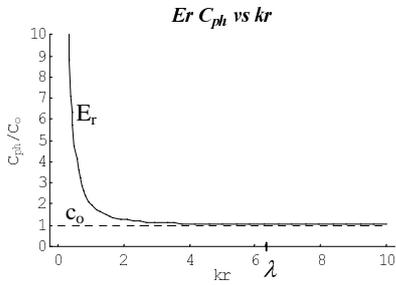

**Figure 6**

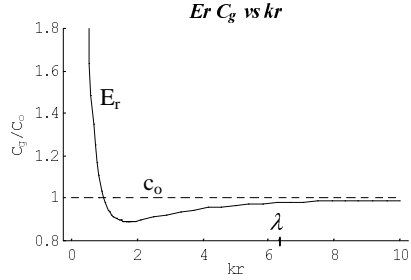

## 3.5 Transverse electric field ($E_\theta$)

Applying the above phase and group speed relations (Eq. 14, 18) to the transverse electrical field ($E_\theta$) component (Eq. 5) yields the following results:

**Figure 7**

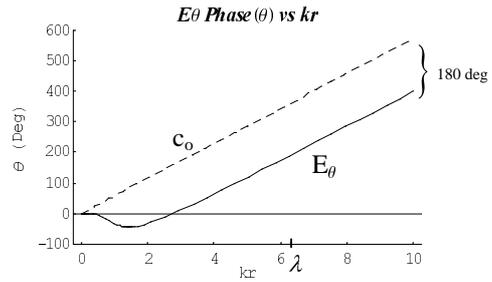

$$y = -kr \qquad x = 1 - (kr)^2 \qquad (23)$$

$$\theta = kr - Cos^{-1}\left(\frac{1-(kr)^2}{\sqrt{1-(kr)^2+(kr)^4}}\right) \qquad (24)$$

$$c_{ph} = c_o\left(\frac{1-(kr)^2+(kr)^4}{-2(kr)^2+(kr)^4}\right) \qquad (25)$$

$$c_g = \frac{c_o\left(1-(kr)^2+(kr)^4\right)^2}{-6(kr)^2+7(kr)^4-(kr)^6+(kr)^8} \qquad (26)$$

**Figure 8**

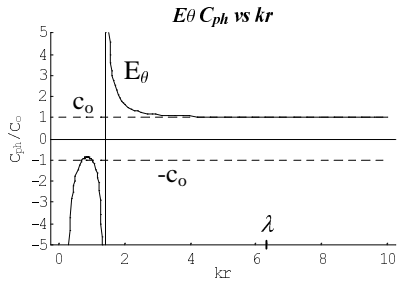

**Figure 9**

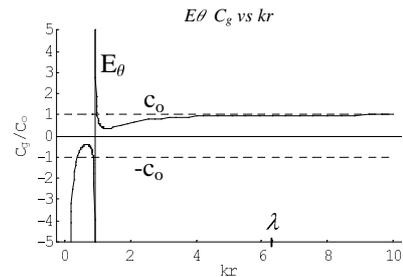

## 3.6 Transverse magnetic field ($H_\phi$)

Applying the above phase and group speed relations (Eq. 14, 18) to the transverse magnetic field ($H_\phi$) component (Eq. 6) yields the following results:

$$y = -1 \qquad x = -kr \qquad (27)$$

$$\theta = kr - \cos^{-1}\left(\frac{-kr}{\sqrt{1+(kr)^2}}\right) \qquad (28)$$

$$c_{ph} = c_o\left(1 + \frac{1}{(kr)^2}\right) \qquad (29)$$

$$c_g = \frac{c_o\left(1+(kr)^2\right)^2}{3(kr)^2 + (kr)^4} \qquad (30)$$

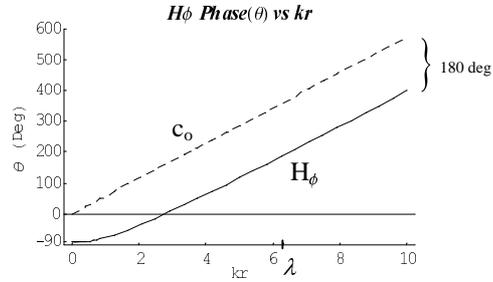

**Figure 10**

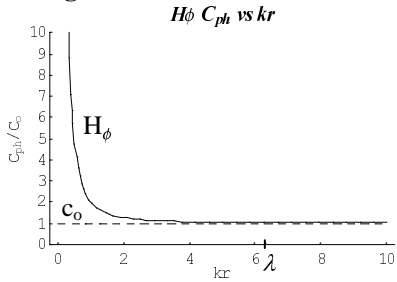

**Figure 11**

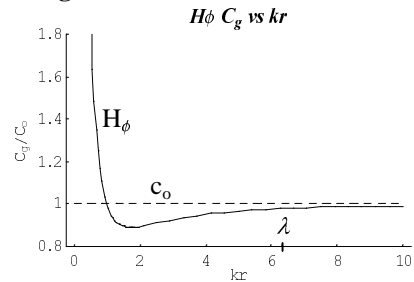

**Figure 12**

## 4  Graphical evidence of superluminal phase and group speed

### 4.1  Superluminal near-field phase speed of radial electric field

To demonstrate the superluminal near-field phase velocity of the longitudinal electric field, the calculated phase and amplitude functions can be inserted into a cosine signal and the field amplitude can then be plotted in the near field as a function of space (r) at several isolated moments in time (t), (Fig. 14). A field propagating at the speed of light (shown as a dashed line) is also included in the plot for reference. The following parameters are used in the subsequent plots: 1m wavelength ($\lambda$), 300GHz signal frequency (f), 3.3ns signal period (T). The following Mathematica code (Fig. 13) is used to generate these plots:

**Figure 13.** Mathematica code used to generate plots

```
L = 1; c = 3*10^8; f = c/L; T = 1/f; w = 2 N[Pi] f; k = w/c;
Animate[Plot[{1/r^2*Sqrt[1/r^2+k^2] *
    Cos[w t - k r + ArcTan[k r]], 20*Cos[w t - k r]},
 {r, 0.1, 3*L}, PlotPoints -> 600, PlotRange -> {-60, 60}],
 {t, 0, 3*T}];
```

**Figure 14.** $E_r$ vs Space – Cosinusoidal Signal

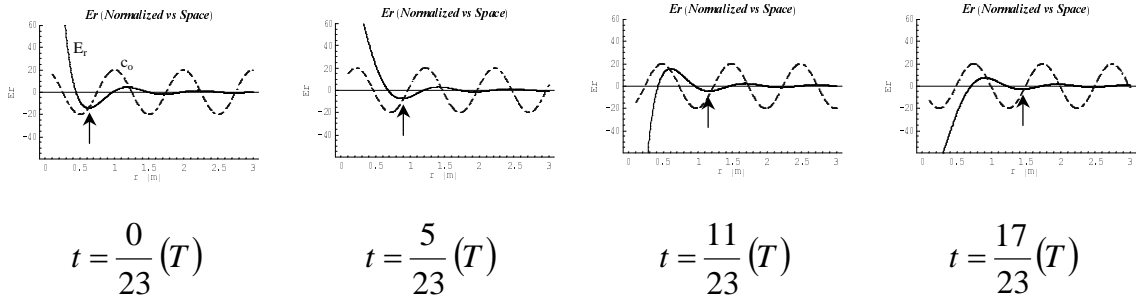

$$t = \frac{0}{23}(T) \qquad t = \frac{5}{23}(T) \qquad t = \frac{11}{23}(T) \qquad t = \frac{17}{23}(T)$$

The longitudinal field (shown as a solid line in the plot above) is observed to propagate away from the source, which is located at r = 0. As it propagates away from the source, the oscillation amplitude decays rapidly ($1/r^3$) near the source ($r < \lambda$), and decreases more slowly ($1/r^2$) in the farfield ($r > \lambda$) (ref Eq. 5). A field propagating at the speed of light (shown as a dashed line in the plot above) is also included in the plot for reference. Both signals start together in phase. The longitudinal field is seen to propagate faster than the light signal initially when it is generated at the source. After propagating about one wavelength the longitudinal electric field is observed to slow down to the speed of light, resulting in a final relative phase difference of 90 degrees. In order to see the effect more clearly the signals can be plotted with the amplitude part of the function set to unity (Fig. 15).

**Figure 15.** $E_r$ (Normalised) vs Space – Cosinusoidal Signal

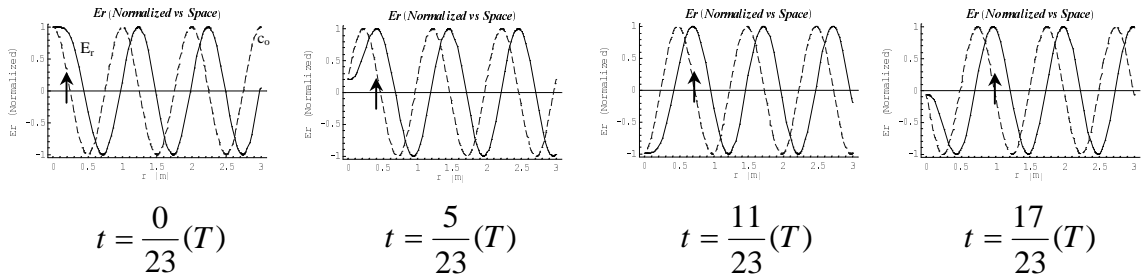

$$t = \frac{0}{23}(T) \qquad t = \frac{5}{23}(T) \qquad t = \frac{11}{23}(T) \qquad t = \frac{17}{23}(T)$$

It is also instructive to plot the signals as a function of time (t) for several positions (r) away from the source (Fig 16). At the source (r = 0) both signals are observed to be in phase. Further away from the source the longitudinal field signal is observed to shift 90 degrees, indicating that it arrives earlier in time. The plots shown below are normalized for clarity, but it should be noted that the signals have the same form even if the amplitude part of the function were included. The only difference is the vertical scaling of the plot. From these plots it can also be seen that the longitudinal field propagates much faster than the speed of light near the source ($r < \lambda$), and reduces to the speed of light at about one wavelength from the source ($r \approx \lambda$), resulting in a final relative phase difference of 90 degrees between the longitudinal field (shown as a solid line), and the field propagating at the speed of light (shown as dashed line).

**Figure 16.** $E_r$ (Normalised) vs Time – Cosinusoidal Signal

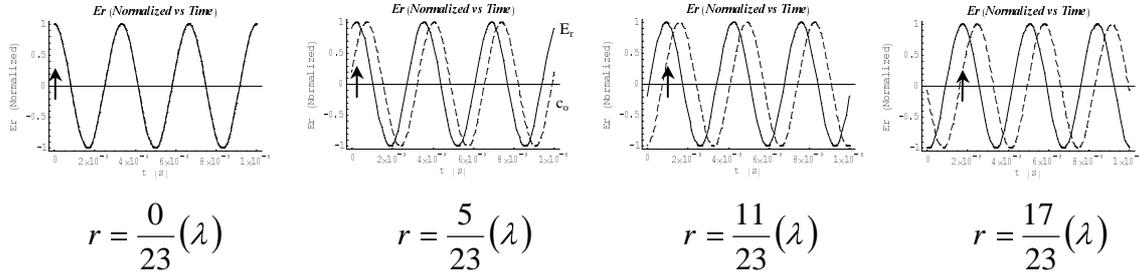

$$r = \frac{0}{23}(\lambda) \qquad r = \frac{5}{23}(\lambda) \qquad r = \frac{11}{23}(\lambda) \qquad r = \frac{17}{23}(\lambda)$$

## *4.2 Superluminal near-field group speed of radial electric field*

To demonstrate the superluminal near-field group propagation speed of the longitudinal field, the calculated phase and amplitude functions can be inserted into the spectral components of an amplitude modulated cosine signal, and the field amplitude can then be plotted as a function of space (r) at several isolated moments in time (t), (Fig. 18). To demonstrate this technique the group propagation (shown as a solid line) is compared to the phase speed propagation (shown as a dashed line) of waves of the form: $Cos(kr - \omega t)$. Note that the phase component (kr) is independent of frequency. This result is known to produce group waves and phase waves that both propagate at the speed of light. This can be seen by using (Eq. 14): since $\theta = kr$ $\therefore c_{ph} = \omega / \frac{\partial \theta}{\partial r} = \omega/k = c_o$. Using (Eq. 18): $c_g = \left[ \partial^2 \theta / \partial r \partial \omega \right]^{-1} = c_o$. The following parameters are used in the following plots: Carrier part of signal - [ 1m wavelength ($\lambda_c$), 300MHz signal frequency (fc), 3.3ns signal period (Tc)] , Modulation part of the signal - [ 10m wavelength, 30MHz signal frequency (fm), 33.3ns signal period (Tm)]. Note that the phase relation of both signals are the same at different isolated moments in time and that both signals propagate away from the source at the same speed. The following Mathematica code (Fig. 17) is used to generate these plots:

**Figure 17.** Mathematica code was used to generate these plots

```
AM = Cos[Wc t] * (1 + Cos[Wm t]); ph1 = kc x; ph2 = (kc - km) x; ph3 = (kc + km) x;
AM1 = TrigReduce[AM]; AM2 = 1/2 * (2 Cos[t Wc - ph1] + Cos[t Wc - t Wm - ph2] + Cos[t Wc + t Wm - ph3]);
L = 1; c = 3 * 10^8; fc = c / L; fm = fc / 10; T = 1 / fc; Wc = 2 N[Pi] fc; Wm = 2 N[Pi] fm;
kc = Wc / c; km = Wm / c; Animate[Plot[{AM2, 2 * Cos[Wc t - kc x]}, {x, 0, 10 * L}], {t, 0, 10 * T}];
```

**Figure 18.** Light Phase (Cosine Wave) and Group (AM Signal) vs Space

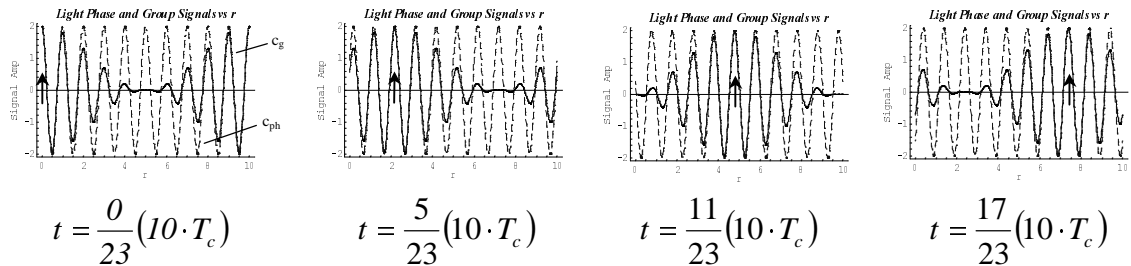

$$t = \frac{0}{23}(10 \cdot T_c) \qquad t = \frac{5}{23}(10 \cdot T_c) \qquad t = \frac{11}{23}(10 \cdot T_c) \qquad t = \frac{17}{23}(10 \cdot T_c)$$

Plotting the signals as a function of time for several spatial positions from the source also shows that group and phase signals travel at the same speed and remain in phase as they propagate.

**Figure 19.** Light Phase (Cosine Wave) and Group (AM Signal) vs Time

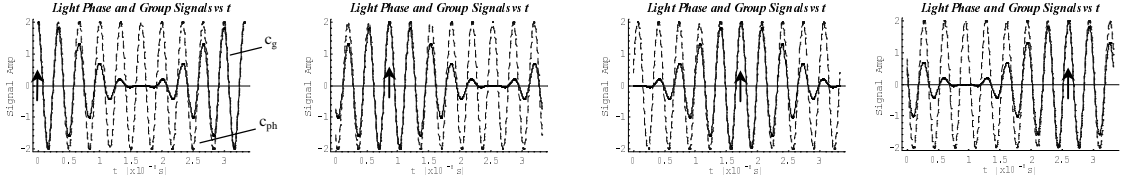

$$r = \frac{0}{23}(10 \cdot \lambda_c) \qquad r = \frac{0}{23}(10 \cdot \lambda_c) \qquad r = \frac{0}{23}(10 \cdot \lambda_c) \qquad r = \frac{0}{23}(10 \cdot \lambda_c)$$

The superluminal near-field group propagation speed of the longitudinal electric field can also be demonstrated in the same way as in the above example. The calculated phase function for the field can be inserted in into the spectral components of an amplitude modulated cosine signal and the field amplitude (shown as a solid line in the plot below) can then be plotted as a function of space (r) at several isolated moments in time (t), (Fig. 21, 22). An amplitude modulated field propagating at the speed of light (shown as a dashed line) is also included in the plot for reference (envelope propagates at speed of light). Note that for this reference signal both the phase speed and the group speed are equal to the speed of light (ref Fig. 18, 19). The following mathematica code (Fig. 20) is used to generate the plots below. The same signal parameters used in the previous example are used in the calculation.

**Figure 20.** Mathematica code used to generate plots

```
AM = Cos[Wc t] * (1 + Cos[Wm t]); ph1 = kc x; ph2 = (kc - km) x; ph3 = (kc + km) x;
ph11 = kc x + ArcTan[-kc x];
ph22 = (kc - km) x + ArcTan[-(kc - km) x]; ph33 = (kc + km) x + ArcTan[-(kc + km) x];
AM1 = TrigReduce[AM]; AM2 = 1/2 * (2 Cos[t Wc - ph1] + Cos[t Wc - t Wm - ph2] + Cos[t Wc + t Wm - ph3]);
AM3 = 1/2 * (2 Cos[t Wc - ph11] + Cos[t Wc - t Wm - ph22] + Cos[t Wc + t Wm - ph33]);
L = 1; c = 3 * 10^8; fc = c / L; fm = fc / 10; T = 1 / fc; Wc = 2 N[Pi] fc; Wm = 2 N[Pi] fm;
kc = Wc / c; km = Wm / c; Animate[Plot[{AM3, AM2}, {x, 0, 10 * L}, PlotPoints -> 600], {t, 0, 10 * T}];
```

**Figure 21.** Er (Normalized) vs Space – AM Signal

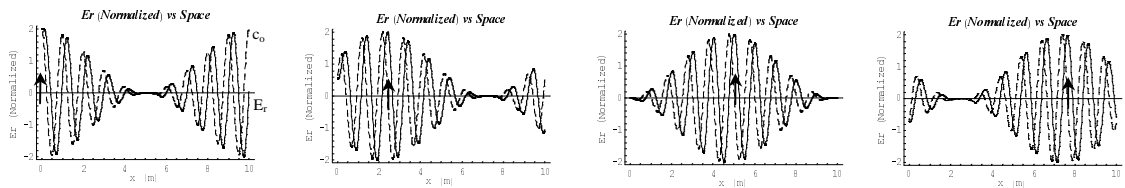

$$t = \frac{0}{23}(10 \cdot T_c) \qquad t = \frac{5}{23}(10 \cdot T_c) \qquad t = \frac{11}{23}(10 \cdot T_c) \qquad t = \frac{17}{23}(10 \cdot T_c)$$

**Figure 22.** Zoom of $E_r$ (Normalized) vs Space – AM Signal

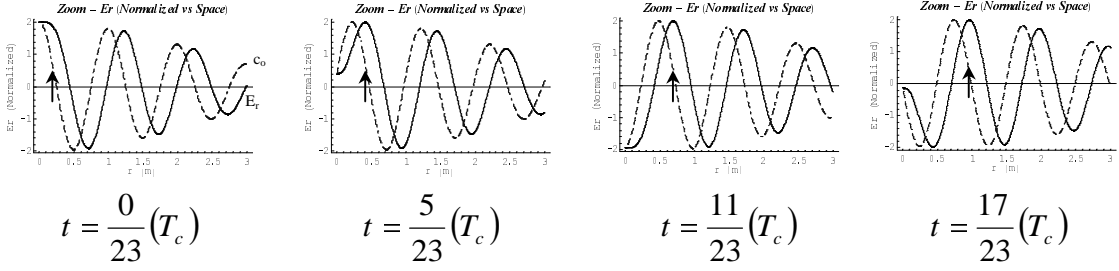

$$t = \frac{0}{23}(T_c) \qquad t = \frac{5}{23}(T_c) \qquad t = \frac{11}{23}(T_c) \qquad t = \frac{17}{23}(T_c)$$

The above plots show an amplitude modulated longitudinal field group packet (shown as a solid line) propagating away from the source, at r = 0 (group maxima marked by vertical arrow). A propagating speed of light group wave (shown as a dashed line) is also provided for reference. The group maxima of the amplitude modulated longitudinal field is observed to propagate to the right side of the plot before the group maxima of the speed of light wave. These series of plots clearly demonstrate that the longitudinal group wave propagates much faster than the speed of light near the source ( $r < \lambda_c$ ). After propagating about one carrier wavelength from the source (r ≈ $\lambda_c$) the modulation part of longitudinal field (envelope of solid line) reduces to the speed of light, resulting in a final relative phase difference of 90 degrees (relative to the carrier signal) between the longitudinal field, and the field propagating at the speed of light.

It is also very instructive to plot the field amplitude of the amplitude modulated longitudinal wave (shown as a solid line in plot below) as a function of time (t), for several positions away from the source (r), (Fig. 23, 24). As before, an amplitude modulated wave traveling with light speed is also plotted for reference (shown as a dashed line). At the source (r = 0) both signals are observed to be in phase. Further away from the source the modulated longitudinal field signal is observed to shift to the left, indicating that the modulation part of longitudinal field (envelope) arrives earlier in time. The plots shown below are normalized for clarity, but it should be noted that the signals have the same form even if the amplitude part of the function were included. The only difference is the vertical scaling of the plot. From these plots it can be seen that the modulation part of longitudinal field (envelope of solid line) propagates much faster than the modulated light speed signal (envelope of dashed line propagates at speed of light) near the source (r < $\lambda_c$). After propagating about one carrier wavelength from the source (r ≈ $\lambda_c$) the modulation part of longitudinal field (envelope of solid line) reduces to the speed of light, resulting in a final relative phase difference of 90 degrees (relative to the carrier signal) between the longitudinal field, and the field propagating at the speed of light.

**Figure 23.** $E_r$ (Normalized) vs Time – AM Signal

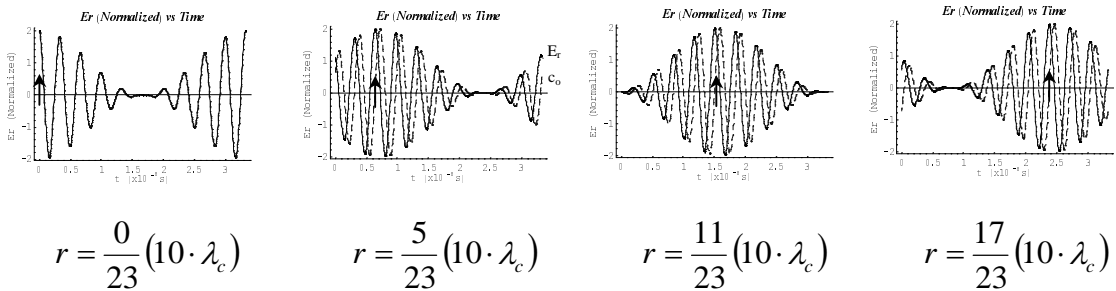

$$r = \frac{0}{23}(10 \cdot \lambda_c) \qquad r = \frac{5}{23}(10 \cdot \lambda_c) \qquad r = \frac{11}{23}(10 \cdot \lambda_c) \qquad r = \frac{17}{23}(10 \cdot \lambda_c)$$

**Figure 24.** Zoom of Er (Normalized) vs Time – AM Signal

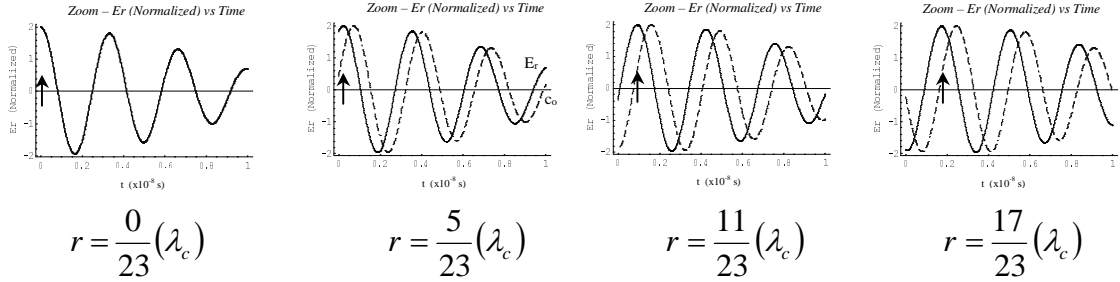

$$r = \frac{0}{23}(\lambda_c) \qquad r = \frac{5}{23}(\lambda_c) \qquad r = \frac{11}{23}(\lambda_c) \qquad r = \frac{17}{23}(\lambda_c)$$

The first frame of the plot shows the two wave groups starting in phase at the source (r = 0). The following frames show that as the two waves propagate away from the source, the group maxima (marked by a vertical arrow) of the amplitude modulated longitudinal arrives earlier in time than the group maxima of a light speed amplitude modulated signal, thus demonstrating that the group speed of an amplitude modulated longitudinal electrical field is much faster than the speed of light near the source (r < $\lambda_c$), and reduces to the speed of light after it has propagated about one wavelength from the source (r ≈ $\lambda_c$), resulting in a final phase difference of 90 degrees.

## 5 Relation between group speed and propagation speed of information and energy

Several authors have indicated that wave group speed (propagation speed of wave envelope) is also the speed at which wave energy and wave information propagates [6](pp.268-269), [10](p.123). One intuitive way to understand this is to mathematically amplitude modulate and demodulate a propagating wave, thereby transmitting and detecting information. An amplitude modulated wave can be mathematically modeled as follows:

$$AM\,Sig = [1 + m\,Cos(\omega_m t)]Cos(\omega_c t) \qquad (31)$$

where ($\omega_m$) is the modulation frequency, ($\omega_c$) is the carrier frequency, and (m) is the index of modulation. Using trigonometric identities the AM signal can be shown to be:

$$AM\,Sig = \frac{m}{2}Cos[(\omega_c - \omega_m)t] + Cos[\omega_c t] + \frac{m}{2}Cos[(\omega_c + \omega_m)t] \qquad (32)$$

Wave propagation can then be modelled by inserting the calculated phase relation of the wave into the spectral components of the modulated signal.

$$AM\,Sig = \frac{m}{2}Cos[(\omega_c - \omega_m)t + \theta_1] + Cos[\omega_c t + \theta_2] + \frac{m}{2}Cos[(\omega_c + \omega_m)t + \theta_3] \qquad (33)$$

If the longitudinal electrical field is used to transmit the signal, then the near-field phase (r < $\lambda_c$), relations for the spectral components are (ref Eq. 20):

$$\text{where } \theta_1 = \frac{(\omega_c - \omega_m)^3 r^3}{3c^3} \qquad \theta_2 = \frac{(\omega_c)^3 r^3}{3c^3} \qquad \theta_3 = \frac{(\omega_c + \omega_m)^3 r^3}{3c^3} \qquad (34)$$

The speed at which the wave group (envelope) propagates can be determined by squaring the resultant modulated signal (Eq. 35) and noting the phase shift ($\theta_{\omega_m}$) of the modulation component ($\omega_m$) of the resultant signal. Performing this computation on the longitudinal electrical field and using trigonometric identities yields:

$$AM\,Sig\,Det = \left[\frac{1}{2}Cos[(\omega_c - \omega_m)t + \theta_1] + Cos[\omega_c t + \theta_2] + \frac{1}{2}Cos[(\omega_c + \omega_m)t + \theta_3]\right]^2$$

$$= \frac{1}{2} + \frac{1}{2}Cos[2\,ph1 - 2\,tWc] + \frac{1}{2}m\,Cos[ph1 - ph2 - tWm] + \frac{1}{2}m\,Cos[ph1 + ph3 - 2\,tWc - tWm] +$$
$$\frac{1}{2}m\,Cos[ph1 - ph3 + tWm] + \frac{1}{2}m\,Cos[ph1 + ph2 - 2\,tWc + tWm] + O[m^2] \quad (35)$$

The resulatant phase shift of the modulation component is: $\theta_{\omega_m} = ph2 - ph3$. Substituting the phase relations (ref Eq. 34) yields:

$$\theta_{\omega_m}\bigg|_{d_o << r \leq \frac{c}{3\omega}} = \frac{-r^3\omega_m(3\omega_c^2 + 3\omega_c\omega_m + \omega_m^2)}{3c^3} \quad (36)$$

The speed at which the modulation envelope (information) propagates can then be calculated using (Eq. 14). Performing this computation on the above result yields the same answer as when calculated using the group speed (ref Eq. 22):

$$c_g = -\omega_m \bigg/ \frac{\partial \theta_m}{\partial r}\bigg|_{d_0 << r \leq \frac{c}{3\omega}} = \frac{c^3}{r^2(3\omega_c^2 + 3\omega_c\omega_m + \omega_m^2)}\bigg|_{\omega_c >> \omega_m} = \frac{c^3}{3r^2\omega_c^2} = \frac{1}{3}c_{ph} \quad (37)$$

Note that this model can also be used to show that the near-field energy density generated by an electric dipole propagates at the group speed, since the energy density (w) of an electromagnetic field is known to be equal to the sum of the squares of each field component[5](p.127):

$$w = \frac{\varepsilon_o}{2}E_r^2 + \frac{\varepsilon_o}{2}E_\theta^2 + \frac{\mu_o}{2}B_\phi^2 \quad (38)$$

## 6 Proposed experiments to measure superluminal near-field phase and group speed of the longitudinal electric field

The previous mathematical arguments have indicated that the group speed of a propagating near-field longitudinal electric field is much faster than the speed of light for propagation distances less than one wavelength (r < λ). The approximate form of the group speed, of the near-field longitudinal electric field (ref Eq. 22) is:

$$c_g = \frac{c_o}{3(kr)^2} = \frac{c_o^3}{3(\omega r)^2} \quad (39)$$

The following experiment is proposed to measure the near-field group speed of a propagating near-field longitudinal electric field. It is suggested that an amplitude modulated signal be injected into one end of a parallel plate capacitor and detected on the other side of the capacitor by an amplitude demodulator. The phase difference between the resultant demodulated signal and the original modulation signal should then be measured. The gap distance of the capacitor plates should then be increased and the phase difference should then be measured again. The group speed can then be determined by entering the value of the modulation frequency (ω), the measured change in phase (Δθ), and the change in capacitor plate gap distance ($d_o$) in the following relation:

$$c_g = \frac{\omega_m d_o}{\Delta\theta} \quad (40)$$

Because the phase speed of the longitudinal electric field is much faster than the speed of light, the expected phase change may not be easy to measure. In the nearfield, at distances less than one tenth wavelength (relative to the carrier frequency), the group speed of the longitudinal electric field is approximately (ref Eq. 22):

$$c_g = \frac{c^3}{3\omega_c^2 d_o^2} \quad (41)$$

$$\therefore \Delta\theta = \frac{\omega_m d_o}{c_g} = 3\frac{\omega_m \omega_c^2 d_o^3}{c^3} \quad (42)$$

Note that using a high modulation frequency and an even higher carrier frequency can increase the observed phase change. If a 50MHz modulation frequency and a 500MHz carrier frequency were used to generate the amplitude modulation signal then a 1mm change in capacitor plate distance would generate a 2x10$^{-5}$ deg phase change. Note that these frequencies correspond to a 6m far-field modulation electrical wavelength and a 0.6m carrier wavelength. This phase change would be very difficult to measure but it may be possible using a high phase sensitivity lock-in technique developed by the author[14].

## 7  Superluminal wave propagation in other physical systems

### 7.1  Electromagnetic fields generated by a magnetic dipole.

The electromagnetic fields generated by an oscillating current loop has been shown by several authors to have a similar form to the electric dipole[1](p261) [3](pp. 623-625, 601). The only difference between the solutions is that electric and magnetic fields are reversed:

Resultant electrical and magnetic fields for an oscillating magnetic dipole

$$H_r = \frac{m_o Cos(\theta)}{4\pi\, r^3}[1 - i(kr)]e^{i(kr)} \qquad H_\theta = \frac{m_o Sin(\theta)}{4\pi r^3}\left[\{1-(kr)^2\} - i(kr)\right]e^{i(kr)} \quad (43)$$

$$E_\varphi = \left(\frac{\mu_o}{\varepsilon_o}\right)^{\frac{1}{2}} \frac{\omega m_o Sin(\theta)}{4\pi r^2}[-kr - i]e^{i(kr)} \quad (44)$$

Consequently, all the analysis performed on the electric dipole in the previous pages can also be applied to this system. Specifically, the near-field phase and group speeds of this system are also concluded to be much faster than the speed of light.

## 7.2  Gravitational fields generated by an oscillating mass

Mathematical analysis of an the gravitational fields generated by a vibrating mass reveals that for weak gravitational fields, along the axis of vibration, the oscillating gravitational dipole is modeled with the a partial differential equation similar to that of the oscillating electric dipole. One significant difference between the two systems is that in order to conserve momentum, a vibrating mass must be accompanied with another oscillating mass oscillating with the same frequency but with opposite phase. The effect of the second vibrating mass is to cancel the propagating gravitational fields in the far field. However, near the source the second vibrating mass does not contribute significantly to the resultant propagating gravitational fields, and can be neglected in the modelling[18].

$$\text{Using the Einstein relation:} \qquad G_{\mu\nu} = \frac{8\pi G}{c^4} T_{\mu\nu} \qquad (45)$$

Along the axis of vibration, for small masses and low velocities, the Einstein equation reduces to:

$$\left( \frac{1}{c_o^2} \frac{\partial^2}{\partial^2 t} - \frac{\partial^2}{\partial r^2} \right) \varphi = \frac{-4\pi G}{c_o^2} T_{oo} \qquad (46)$$

The only non-vanishing term in the energy momentum tensor to order $\beta^2$ is:

$$T_{oo} = \rho c^2 \delta[r - d_o Sin(\omega t)][1 + O(\beta^2)] \qquad \text{in which:} \quad \beta = \frac{\omega d_o}{c_o} \qquad (47)$$

Solving the partial differential equation yields[17]:

$$\varphi = \frac{K_o}{r} \left[ \frac{1}{1 - \xi Sin\left(\omega t - \frac{\omega l'}{c_o}\right)} \frac{1}{1 - \beta Cos\left(\omega t - \frac{\omega l'}{c_o}\right)} \right] [1 + O(\beta^2)] \qquad (48)$$

$$\text{in which:} \qquad K_o = -mG, \qquad \xi = \frac{d_o}{r}, \qquad l' = r - d_o Sin\left(\omega t - \frac{\omega l'}{c_o}\right) \qquad (49)$$

The gravitational field (g) can then be calculated using the relation $\quad g = -\frac{\partial}{\partial r}\varphi$:

$$g = \frac{-mG}{r^2} \cdot \left[ 1 - \left(\frac{d_o}{r}\right)^2 \right]^{\frac{-3}{2}} \left[ 1 + 2\frac{d_o}{r} Sin\left(\omega t + O(\beta)^2\right) \right] [1 + O(\beta)^2] + h.h. \qquad (50)$$

Since $\theta = \dfrac{\omega r}{c_{ph}} = O\left[\left(\dfrac{\omega d_o}{c_o}\right)^2\right]$, solving this relation for the phase speed of the longitudinal gravitational field yields:

$$\dfrac{c_{ph}}{c_o} \geq O\left[\left(\dfrac{r}{\omega d_o^{\,2}}\right)c_o\right] \qquad (51)$$

Therefore it is concluded that near the source, along the axis of vibration, the longitudinal gravitational phase speed is much faster than the speed of light. The group velocity has been shown to be

$c_g = \left[\dfrac{\partial^2 \theta}{\partial r \partial \omega}\right]^{-1} = \left[\dfrac{\partial}{\partial \omega}\left(\dfrac{\partial \theta}{\partial r}\right)\right]^{-1}$ (ref Eq. 18). Inserting the relation for the phase speed (ref Eq. 14) yields: $c_g = \left[\dfrac{\partial}{\partial \omega}\left(\dfrac{\omega}{c_{ph}}\right)\right]^{-1}$. Inserting the order estimate of the phase speed (Eq. 51) yields:

$$\dfrac{c_g}{c_o} \geq O\left[\left(\dfrac{r}{2\omega d_o^{\,2}}\right)c_o\right] \qquad (52)$$

From the above results (Eq. 51, 52) it is concluded that near the source both the phase speed and the group speed of the longitudinal gravitational field, along the axis of vibration of the mass, are much faster than the speed of light.

# 8  Conclusion

This paper has provided mathematical evidence that electromagnetic near-field waves and wave groups, generated by an oscillating electric dipole, propagate much faster than the speed of light as they are generated near the source, and reduce to the speed of light at about one wavelength from the source. The speed at which wave groups propagate (group speed) has been shown to be the speed at which both the wave energy density and modulated wave information propagate. Because of the similarity of the governing partial differential equations, two other physical systems (magnetic oscillating dipole, and gravitational radiating oscillating mass) have been shown to have similar near-field superluminal results.